# 2D-MoS$_2$ with Narrowest Excitonic Linewidths Grown by Flow-Less Direct Heating of Bulk Powders

*Davoud Hejazi[1], Renda Tan[1], Neda Kari Rezapour[2], Mehrnaz Mojtabavi[2], Meni Wanunu[1,2], and Swastik Kar[1]*

[1]Department of Physics and [2]Department of Bioengineering, Northeastern University, Boston, MA 02115, USA

## ABSTRACT

Developing techniques for high-quality synthesis of mono and few-layered 2D materials with lowered complexity and cost continues to remain an important goal, both for accelerating fundamental research and for applications development. We present the simplest conceivable technique to synthesize micrometer-scale single-crystal triangular monolayers of MoS$_2$, *i.e.* by direct heating of bulk MoS$_2$ powder onto proximally-placed substrates. Room-temperature excitonic linewidth values of our samples are narrower and more uniform than those of 2D-MoS$_2$ obtained by most other techniques reported in literature, and comparable to those of ultraflat h-BN-capped mechanically exfoliated samples, indicative of their high quality. Feature-rich Raman spectra absent in samples grown or obtained by most other techniques, also stand out as a testament of the high quality of our samples. A contact-growth mode facilitates direct growth of crystallographically-strained circular samples, which allows us to directly compare the optoelectronic properties of flat *vs.* strained growth from the same growth runs. Our method allows, for the first time, to quantitatively compare the impact of strain on excitonic and Raman peak positions on identically-synthesized 2D-MoS$_2$. Strain leads to average Red-shifts of ~ 30 meV in the A-exciton position, and ~ 2 cm$^{-1}$ in the E$^1_{2g}$ Raman peak in these samples. Our findings open-up several new possibilities that expand 2D material research. By eliminating the need for carrier gas flow, mechanical motion or chemical reactions, our method can be either miniaturized for substantially low-cost, high-quality scientific research or potentially scaled-up for mass-production of 2D crystals for commercial purposes. Moreover, we believe this technique can also be extended to other transition metal dichalcogenides and other layered materials.

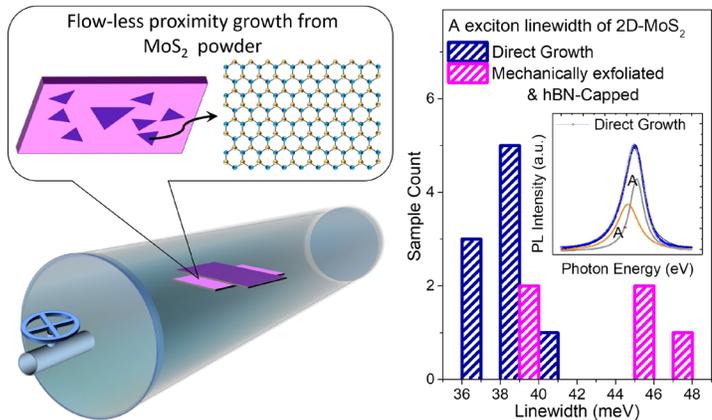

Author for correspondence: Swastik Kar
Email: s.kar@northeastern.edu

## Introduction

Two-dimensional transition metal dichalcogenides (2D-TMDs) are beyond-graphene layered materials that have become the new platform for studying the physics of 2D semiconductors.[1–5] With an atomically-thin layers confined in a 2D plane, 2D-TMDs manifest remarkable properties including indirect-to-direct bandgap switching,[6–8] emergent photoluminescence,[9] strong photovoltaic response,[6,9] anomalous lattice vibrations,[10] strong light-matter interactions at heterojunctions,[11–14] valley-selective circular dichroism,[15–17] excitonic dark states,[18–20] control of valley polarization using optical helicity,[21,22] and field-induced transport with a current ON-OFF ratio exceeding $10^8$,[23,24] that gives 2D-TMDs immense potential for transistors, photodetectors, sensors, many other applications.[25–42] Among the variety of materials being investigated, the thinnest semiconductor,[25] molybdenum disulfide (2D-$MoS_2$), exhibits promising prospects for low-cost, highly sensitive, and flexible next-generation optoelectronics, nanoelectronics, photovoltaics, and valleytronics applications.[43–50] Unlike graphene that does not manifest a bandgap,[51] 2D-$MoS_2$ has layer thickness-dependent bandgap,[52] which is indirect in bilayer and above but becomes direct in the monolayer limit.[53] It has also been shown that it is possible to obtain the valley polarization of excitons using circularly polarized light excitations.[54–56] Moreover, the sheet resistance of 2D-$MoS_2$ can be easily controlled either by applying a gate voltage, incident light, or injecting concentrations of dopants.[57–59] Further, 2D-$MoS_2$ is a strongly interacting system even in the presence of relatively high carrier densities.[60,61] These properties turn 2D-$MoS_2$ into an ideal laboratory for exploring many-body phenomena, and a highly tunable and prime candidate for a wide range of applications such as photoemitters, phototransistors, and photodetectors.[62–73]

Excited-state dynamics in monolayer TMDs is sensitive to their quality, and their relaxation pathways are affected both intrinsic (*e.g.* e-e, e-phonon interactions) and extrinsic (*e.g.* defect, temperature *etc.*) factors.[74–78] Hence, investigating photoexcited processes help us compare the quality of 2D materials. The quasiparticle bandgap ($E_g$ ~ 2.4 eV in monolayer $MoS_2$)[52,79] characterizes single-particle (or quasiparticle) excitations and is defined by the sum of the energies needed to separately inject an electron and a hole into monolayer TMD.[6,80–83] The optical bandgap ($E_{opt}$ ~ 1.85 eV in monolayer $MoS_2$) describes the energy required to create an exciton in its ground state, a correlated two-particle electron-hole pair, via optical absorption.[52,84,85] The difference in these energies ($E_g - E_{opt}$) directly yields the exciton binding energy ($E_b$),[6,86] which is about 20 times that of kT ~ 25 meV at room temperature for in monolayer $MoS_2$, hence excitons are called tightly bound in 2D materials.[84,85] In TMDs, enhanced Coulomb interactions due to low-dimensional effects are expected to increase the quasiparticle bandgap as well as causing electron-hole pairs to form more strongly bound excitons.[83,87,88] Photoluminescence (PL) measurements in charge-neutral 2D-$MoS_2$ show two excitonic peaks, associated with A-excitons and B excitons, each originating from one branch of the spin-orbit-split valence-bands near the K-points of its first Brillouin zone.[54] Typically, substrate-induced injection of electrons lead to n-type doping of monolayer $MoS_2$, and results in formation of stable trions, $A^-$, with a slightly lower peak position.[89–91] The sharpness of the PL linewidths associated with each of these excitonic peaks *i.e.*, the full width at half maximum (FWHM) of excitonic/trionic peaks, is accepted as a non-perturbative measure of the quality of the 2D semiconductor,[92–95] since the linewidth in energy scale is inversely proportional to the lifetime of the excitation, *i.e.*, how long it takes for exciton and/or trion to recombine.[54,87] The linewidth is also an indicator of homogeneity/inhomogeneity of the material, *i.e.*, whether it is a single crystal and shows uniform electronic/optoelectronic responses.[60] In an ideal situation where the material is homogenous, and all transitions are direct, the lineshape is expected to be narrow and obey a Lorentzian distribution;[96] as inhomogeneity and lattice vibrations *i.e.*, phonons, increase, there are additional contributions from indirect transitions as well, and the lineshape starts to become broader and follow a Gaussian pattern.[96–99] However, it is worth mentioning that the crystal quality-dependent change in lineshape is different from the temperature-dependent linewidth change. At low temperatures, the PL is expected to be narrow, and by approaching the absolute zero kelvin, the PL lineshape theoretically should approach the Dirac delta function.[100,101] At room temperature, linewidth widening is also an effect of the temperature rising above absolute zero, which according to Fermi-Dirac distribution, results in the change in Fermi function and, in turn, causes the increase in the linewidth of the exciton.[60,89,102,103] Thermal effects such as exciton-phonon coupling and density of states, also doping concentrations can change the overall lineshape of the PL, not merely the linewidth.[60,99,100] Although there are multiple factors affecting the PL line-shape and linewidth, as long as the thermal effects, doping, and other factors are assumed to be the same, the only remaining factor that affects the linewidth is how well the 2D sample is synthesized, or, in other words, how disordered the crystal is. For this reason, the linewidth is a good measure of the quality of the 2D-MoS$_2$. We note that it is common to use mobility as a measure of 2D material quality.[104–109] While mobility is clearly an important parameter for quantifying the quality of 2D material, its measurement requires subjecting the sample to lithographic steps which introduces unavoidable chemical contamination,[110] possible contact-resistance-limitations,[111] and accurate estimations of sample geometry. In comparison, optical measures such as PL and Raman can be performed on as-grown crystals without any modifications, and hence we use this as a better measure of the quality of the pristine samples.[112–115]

Obtaining high-quality 2D-TMDs that represent suitable properties both for enabling the demonstration of sensitive quantum phenomena, as well as for various applications, especially for high-performance optoelectronics, has so far been limited by the synthesis techniques.[116–118] It is believed that the highest quality 2D samples, characterized by their narrow photoluminescence (PL) linewidth, can only be obtained by the mechanical exfoliation (ME) of the atomic layers of TMDs from their bulk crystals.[119–124] Field-effect transistors (FETs) made from post-processed ME samples have high ON-OFF switching ratios, high field-effect mobilities, and are sensitive to certain ranges of the visible spectrum.[125] However, there are significant challenges associated with ME in their inefficiency and difficulty of large-scale production, small lateral sample sizes, and spatial non-uniformities. Moreover, in order for any 2D samples to exhibit their high-quality properties, one has to make them extremely flat, which is only possible by capping them with Boron-Nitride (h-BN).[126,127] The best known 2D-TMDs are hBN-capped ME samples, so the capping step adds to the complications of obtaining high-quality flat 2D samples. Hence, even though the ME technique for obtaining high-quality 2D samples is attractive, its poor yield,[128] uncontrollable and irregular sample homogeneity, and not being scalable makes this technique unsuitable for almost any practical applications.[128,129] Chemical vapor deposition (CVD), on the other hand, is a scalable technique, where, unlike ME, large-scale single crystals of 2D-TMDs with uniform layer-thicknesses over lateral sizes reaching hundreds of micrometers can be produced.[52,130] In CVD, TMDs are typically grown using $MoO_3$ and X (X=S, Se, W *etc.*) as the precursors, and the samples are synthesized through a multi-step chemical reaction of one or more precursors, usually in an inert atmosphere, where there is a flow of one or more carrier gases, and detailed control of temperature, pressure, flow-rate, precursors-substrate distance, precursor-precursor distance, the temperature at each precursors location as well as at substrate location, *etc.* are crucial for high-quality homogenous growth. CVD-produced 2D-TMDs are regarded as the high-potential candidates for practical industry-level integration with current CMOS platforms,[131–133] but are still known to be of poor optoelectronic quality and poor yield, which has its root in the probabilistic nature of its two-step chemical deposition process.[52,134] In our previous work, we introduced vapor phase chalcogenization (VPC), a one-step chemical reaction process that results in optoelectronic-grade 2D-TMDs.[52,135] In this method, the direct chemical conversion of $MoO_2$ to $MoS_2$ or $MoSe_2$, results in more complete crystalline conversion into the 2D-TMD samples, even without post-treatment, and hence were comparable to the ME samples, making VPC a suitable technique for practical applications. This suggests that simplifying the number of chemical steps contributes towards increase crystallographic quality.

Wu *et al.* takes this approach a step further, and show that it is possible to grow 2D-$MoS_2$ by flowing Argon gas over heated bulk $MoS_2$ powder and allowing them to condense downstream on insulating substrates such as $SiO_2$ and sapphire, where the crystallographic quality was indirectly established by demonstrating valley polarization.[136] However, the direct comparison of these 2D crystals with those produced by other means was not established. Moreover, this method still involves space-occupying components such as quartz tubes, furnaces, flow-controllers, gas tanks and associated flow-lines and valves. Additionally, this technique also required a detailed control of precursor amounts, and their distances from each other as well as from substrate. As it is well known, the presence of so many variables multiply the uncertainly for obtaining high-quality, reproducible samples. In other words, a simple fabrication technique without the need for multiple control parameters would be far more attractive for advancing the science and applications of 2D materials.

In this report, we present the simplest conceivable technique for synthesizing 2D materials till date, *i.e.* the flow-less direct growth (DG) of 2D-$MoS_2$ by heating commercially-purchased bulk $MoS_2$ powder from a source onto proximally-placed substrates, kept in an Argon atmosphere. The chemical-reaction-free transformation from bulk to vapor to 2D morphology suggests that formation of mono and few-layers was thermodynamically the most preferred morphology, and the absence of any oxygen and carrier-gas flow, as well as the physical proximity of the substrate substantially eliminated the possibility of oxidation during crystal growth. As a result, the 2D-$MoS_2$ samples fabricated by our technique possess some of the narrowest room-temperature excitonic linewidths reported in literature to date, with the best A-exciton linewidth values as low as ~ 36 meV. This is much lower compared to bare ME samples, and comparable to those of h-BN-capped ME samples which are known to have the narrowest achievable linewidths. The average A-exciton linewidth from our samples is ~ 40 meV with a standard deviation of 2.94 meV (*i.e.* <10% standard deviation in quality over several synthesis runs), which reflects extreme homogeneity for any "grown" 2D materials. Our method overcomes the persisting complications such as the need for multiple precursors and carrier gases, and hence paves the way for on-demand miniaturization of 2D-TMD synthesis. Unlike past reports, no substrate pre-treatment, sample post-treatment such as capping, or *in situ* annealing is required in this technique in order to achieve samples with high-qualities comparable to post-processed ME samples. The directly grown samples by our DG technique manifest high optical responses, which is evident in their strong PL and feature-rich Raman spectra; see the Results and Discussion section. The results of this research, as well as our previous work, suggest that at least for TMDs, synthesized monolayers can be comparable if not better than mechanically-exfoliated samples. Interestingly, where the growth substrate is not in contact with the source ($MoS_2$ powder), uniform, triangular single-crystal 2D-$MoS_2$ grows, as expected from the hexagonal lattice structure of $MoS_2$. However, at places where the substrate touches the source powder, the $MoS_2$ nucleation sites create spatial constraints on the formation of the crystals, where we see "strained" samples, forced to "wrinkle" and form strained layers that appear in circular symmetries around the nucleation sites where the substrate has been in contact with the source. This phenomenon enables us to control the strain-induced defects, especially wrinkles, in 2D materials which is gaining attention as a attractive method for inducing novel phenomena[137] and applications.[138–140] In the next sections, we present details of the synthesis, characterizations, and analysis of the quality of samples obtained in comparison to those from a variety of existing synthesis techniques.

**Results and Discussion**

In Figure 1 (a), we depict our flow-less direct growth technique's schematics, where a silicon wafer substrate with a 300 nm thick silicon dioxide ($Si/SiO_2$) coating is placed facing down to the source, and together are placed inside a small quartz tube. The air is pumped out, the tube is back-filled with an inert gas, Argon (Ar) in this case, and is sealed. However the excess pressure was allowed to release through the valve at high temperature, and the growth pressure is kept around 800 Torr, a bit above the atmospheric pressure, to prevent the air re-entering the tube, but there is no carrier gas flow; see Methods for details. Since the as-purchased $MoS_2$ powder has large grains, we broke the powder grains into smaller, particles by ultrasonication in isopropyl alcohol (IPA), and uniformly drop-casted the resulting suspension onto one $Si/SiO_2$ chip as the source.

In Figure 1(b), we show schematics of the growth conditions which enable controlled growth of flat or strained 2D-$MoS_2$. It is not surprising to obtain flat *i.e.,* triangular samples when there is enough distance between powder and substrate, and vapor-transport is the only means of growth; however, as the schematics for the strained samples illustrate, we found that strained samples can controllably grow around a central physical seed *i.e.*, where the powder comes in contact with the source. In this situation, 2D materials have to conform to the strain enforced by the physical seed; thus, samples start to grow in circular patterns, the phenomenon that wrinkles and even fractures the 2D samples. These findings are exciting in the study of deformed 2D-TMD crystals and applications that require defected 2D-TMDs. Figure 1 (c) shows typical triangular samples, and Figure 1 (d) shows typical circular (strained) samples, where we see layers of 2D-MoS$_2$ being strained around the central physical seed. We also see triangular second-layer crystals grown on top of the circular regions.

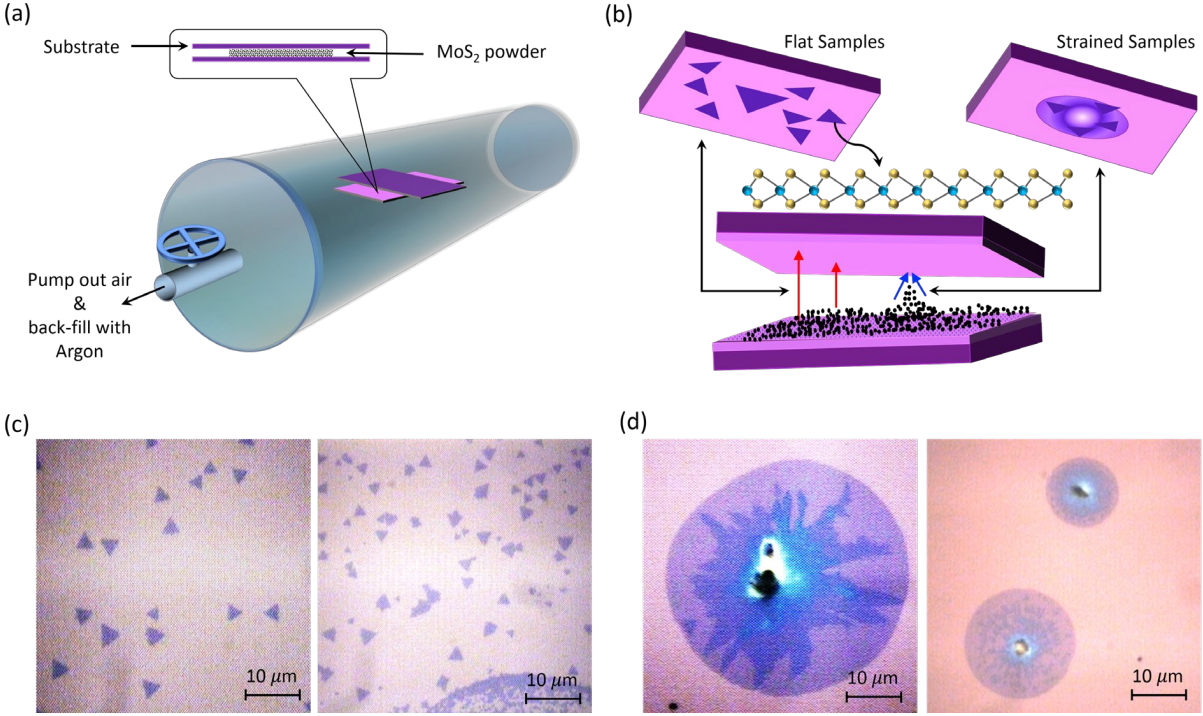

***Figure 1**(a). Schematic setup for direct growth of 2D-MoS$_2$. Powdered bulk MoS$_2$ is evenly spread out on an Si/SiO$_2$ chip (source), and then the substrate, a second Si/SiO$_2$ chip is placed facing down, directly on the powder. These are placed inside a quartz tube from which the air is pumped out and back-filled with Argon (several times) and valve-sealed. The gas source is then disconnected, and there is no further carrier gas flow. The sealed tube is heated up to the growth temperature of 750 °C and kept for 40 min (see text for details). (b) Schematic depiction of the source, substrate and the two possible growth mechanisms. When the source powder is not in contact with the substrate (red arrows), 2D-MoS$_2$ grows in the form of flat triangles, whereas if the powder comes in contact with the substrate (blue arrows), 2D-MoS$_2$ is forced to grow in the form of strained circle-like patterns around the contact sites (i.e., nucleation sites). The side view of MoS$_2$ crystal is also given, which shows one atomic layer of its lattice is about three atoms thick. (c) Optical images of triangular 2D-MoS$_2$ crystals that are grown by the direct method. (d) Optical images of strained circular 2D-MoS$_2$ samples that are grown by the direct method.*

For a more detailed study of the surface topology of these directly grown samples, we captured their atomic force microscopy (AFM) images; the results are shown in Figure 2, where there are three AFM images and two cross-section line profile plots for each image. We can see a typical triangular sample in Figure 2 (a) and its two cross-section line profiles in Figure 2 (b) and (c), estimating the step-height of the edge of the 2D-MoS$_2$ on Si/SiO$_2$ to be ~ 1 nm. Figure 2 (d) shows an AFM image from one portion of a strained circular sample, where we can clearly see the stacked layers, as well as wrinkles on the sample. This figure's inset shows the optical image of the sample's location from where the AFM image is acquired. As we can see in Figure 2 (e), the step-height of the edge of the circular sample is ~ 0.9 nm, almost the same as the triangular sample; but we also see a triangular crystal appearing on top of the circular part, which has the step-height of ~ 0.65 nm. Figure 2 (f) reveals the aforementioned wrinkles on the strained samples; as we move to the center of the strained sample, the wrinkles become more prominent, which is expected since it is approaching the center of constraining geometry. We can see the same pattern in the last three figures acquired from a different part of the strained sample. To eliminate the possibility that these circular samples grow from possible defect sites on the substrate (as against our proposed contact-induced seeding), we note that these structures were never found to grow in "non-contact" mode – *i.e.* either when the substrate was physically separated from the source powder, or during VPC synthesis. Additionally, each circular patch is characterized by a tall hillock at its center, providing evidence that the center was directly in contact with the MoS$_2$ bulk powder at the time of growth. We note that in some cases, microns-scale MoS$_2$ particles may have been electrostatically transferred from the source to the substrate during growth due to the proximity of the substrate from the source-powder surface, thus forming the seed for these circular samples. Growth from such a site can be expected to be similar to that of contact-induced seeding, the resulting growth mechanism can be expected to be similarly strained.

To study the optical properties, exciton/trion linewidth, and vibrational modes of these samples, we excited the 2D-MoS$_2$ samples by a 488 nm laser. We found that in general the highest intensity values of PL were obtained from the exterior parts of the circular samples, which were mostly strained monolayer regions susceptible to bandgap modulation that is known to enhance PL.[141] The normalized PL *vs.* photon energy of three types of samples are given in Figure 3(a), taken from a typical directly grown (DG)-triangular sample, a typical directly-grown (DG)-circular strained sample, and a typical VPC-grown sample for comparison. As we discuss later, each obtained PL spectrum was Lorentz-curve-fitted to obtain the relative positions and contributions of excitons and trions. The Inset of Figure 3(a) shows the average values of A-exciton peak positions over all of the samples of the same type and their standard deviations (11 DG-triangular, 11 DG-circular, and 15 VPC samples), revealing two interesting observations. First, the A-exciton peak of directly-grown triangular samples manifests higher peak energies on average, with very little deviation – suggesting the high uniformity of sample quality for this type of samples. In comparison, the A-exciton peak for both DG-circular and VPC-grown samples have lower peak energy positions and wider spreads (larger standard deviations). While spread-out values are expected in the strained samples due to the random nature of the strain/wrinkles in these sample, the much lower spread in the DG-triangular samples compared to VPC-grown samples suggested that DG-triangular architectures are far more uniform than the latter as well. Our approach allows for the first time to compare the A-exciton position between strained and un-strained (flat) samples grown in the same run. We find that strain in our samples led to average red-shifts of ~ 30 meV in the A-exciton peak position. Second, the similarity of their peak position and variation also suggests that VPC-samples may have larger intrinsic strain compared to the DG-triangular samples. The similarity between the DG-circular and VPC samples is also reflected in Raman peak positions. Figure 3(b) shows the Raman spectrum *vs.* wavenumber of the same three types of samples, with the signature $E^1_{2g}$ and $A_{1g}$ Raman modes for $MoS_2$.[142–144] These graphs were smoothened and normalized with respect to silicon peaks (from the substrate) that appear at 520 cm$^{-1}$. The Inset of Figure 3(b) shows the average Raman peak separations between $A_{1g}$ and $E^1_{2g}$ (*i.e.* $\Delta = \omega [A_{1g}] - \omega [E^1_{2g}]$) – the value of which is expected to be between ~ 18-22 cm$^{-1}$ for monolayer $MoS_2$)[52] – collected from all of the samples of the same type, and their standard deviations. In this case, we find $\Delta$ is smaller for the DG-triangular samples compared to the strained or VPC samples – suggesting that increasing strain within the crystal is at least partially responsible for the higher values of $\Delta$. We were also able to quantify the impact of strain on our Raman peak positions. Strain led to average red-shifts of ~ 2 cm$^{-1}$ in the $E^1_{2g}$ Raman peak position in our samples. From these results, it appears that the DG-triangular samples to have both higher crystallinity and lower intrinsic strain compared to VPC-grown samples. Finally, Figure 3(c) shows the as-collected Raman spectra that appear in Figure 3(b), but significantly magnified to reveal prominent Raman-active modes in $MoS_2$. In our previous reports, we had established that optoelectronic-grade VPC-grown TMDs appear to reveal significantly higher number of Raman peaks as compared to those from other methods.[52,145] In this figure, we find that DG-grown samples reproduce every single one of those rich Raman modes of 2D-$MoS_2$ attributed to the various lattice vibrational modes of this material under optical excitation; see Bilgin *et al.* for comparison.[52] Taken together, our Raman spectral analysis also confirms the high crystalline quality of DG-grown samples. We next investigate in detail the lineshape analysis of PL spectra, which can be considered to be one of the most stringent tests to evaluate the crystalline quality of TMDs.

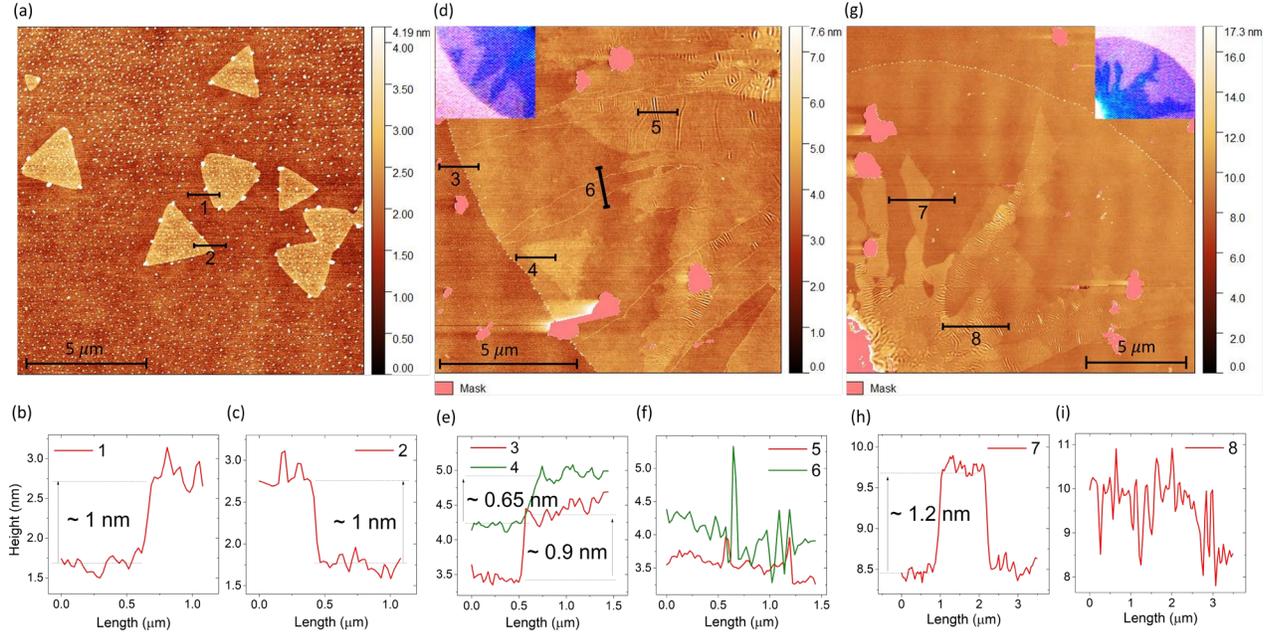

***Figure 2**(a). Atomic force microscopy (AFM) image of a directly grown triangular sample on Si/SiO$_2$ substrate. (b) and (c) The two chosen areas' cross-section line profiles are shown in the previous AFM image. (d) The AFM image of a strained circular sample. (e) and (f) The cross-section line profiles of the chosen areas. (g) The AFM image of a different part of the strained circular sample. (h) and (i) The cross-section line profiles of the chosen areas. The AFM images' insets are the optical images taken from the same locations where AFM images are obtained.*

Lattice vibrations and lattice imperfections affect the lineshape of the PL spectrum.[146–148] When the coupling between the exciton and lattice vibrations or phonons is sufficiently weak, the lineshape is expected to be Lorentzian, which is often used to curve-fit the exciton and trion in the literature.[96] For this reason, we used

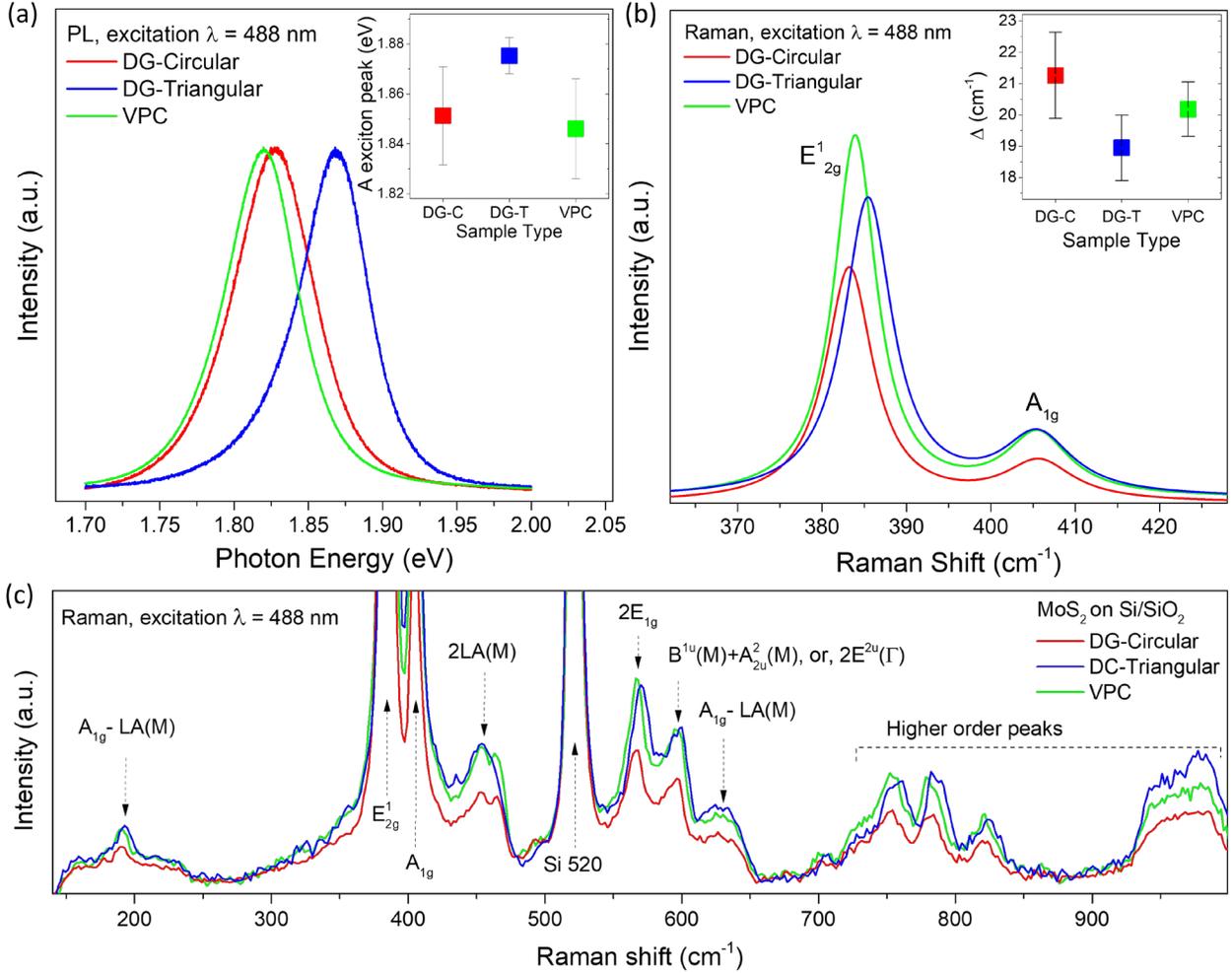

*Figure 3(a). Normalized photoluminescence (PL) spectra of a typical directly grown triangular sample, a directly grown strained circular sample, and a VPC-grown sample as a comparison. The inset figure shows the average A-exciton peak position over all the same type samples and their standard deviations. (b) Smoothed Raman spectra vs. wavenumber of the same three types of samples, normalized with respect to their respective Si peak that appear at 520 cm$^{-1}$. The inset shows average $\Delta = \omega[A_{1g}] - \omega[E^1_{2g}]$ over all the same type samples, and their standard deviations. (c) Magnified Raman spectra that appear in the previous figure. Here, the detailed Raman modes of 2D-MoS$_2$ are shown, attributed to the various lattice vibrational modes of this material under a 488 nm excitation.*

Lorentzian functions to fit A-exciton and A$^-$-trion in the PL data (we note that the B-exciton peak was strongly suppressed in our sample). Figure 4(a) shows typical curve-fits to these samples, where the excitons and trions are labeled. For comparison, we also performed an extensive analysis on the linewidth of 2D-MoS$_2$ samples fabricated on Si/SiO$_2$ by various other techniques reported in literature. Analyses were performed either using the published numerical data in articles or by digitizing the PL data from the published images within these articles. Curve-fitting was performed using Lorentzian functions in the same way we did for our samples. The histogram of A-exciton linewidths of the 2D-$MoS_2$ samples fabricated by various techniques are shown in Figure 4 (b), and the histogram of corresponding $A^-$-trion linewidths are shown in Figure 4 (c). We find that that the median A-exciton line-width of DG-triangular, DG-circular, VPC-grown, un-treated ME and CVD-grown samples are 39.37, 41.44, 62.17, and 64.24 meV, respectively. Taking linewidth narrowness as our comparison metric, we find the remarkable result that from Figures (b) and (c) that the directly grown and VPC samples are among the best quality as-grown/fabricated samples. Further, with the median linewidth of h-BN-capped ME samples at 40.92 meV, similar to that of the directly-grown triangular samples – suggesting that our directly-grown triangular samples are intrinsically superior in quality when compared to some of the best samples reported in literature. The relatively more compact distribution of the linewidths for DG and VPC technique grown samples suggest that samples fabricated through these approaches are uniformly of higher quality, compared to many other techniques whose linewidth distributions are far more spread-out. As expected, CVD-grown samples also reveal the lowest quality and somewhat random probability of getting relatively good samples. A similar comparison of the A-trion linewidths reveals a similar picture, *i.e.* directly grown and VPC samples appear to have comparable linewidths as ME (both h-BN-capped and uncapped) samples which are far superior to that of CVD-grown samples, and with much higher homogeneity of linewidth distributions.

Finally, we perform a more stringent comparison between our DG samples with ME samples. The underlying mechanisms that govern the overall PL and consequently, exciton/trion lineshape is a much-debated subject.[112,149–151] As mentioned earlier, in an ideal situation, the PL is a sum of Lorentzian distributions; However, as lattice vibrations and defects start to perturb the exciton-phonon coupling, it adds a Gaussian components to the statistical distribution as well. Even though curve-fitting the PL to a set of Lorentzian functions is an accepted method by the majority of the TMD community, some researchers also use a combination of Lorentzian and Gaussian fitting functions.[152–156] This distribution does not have a closed-form solution and must be solved via numerical approaches, using the so-called "Voigt" function.

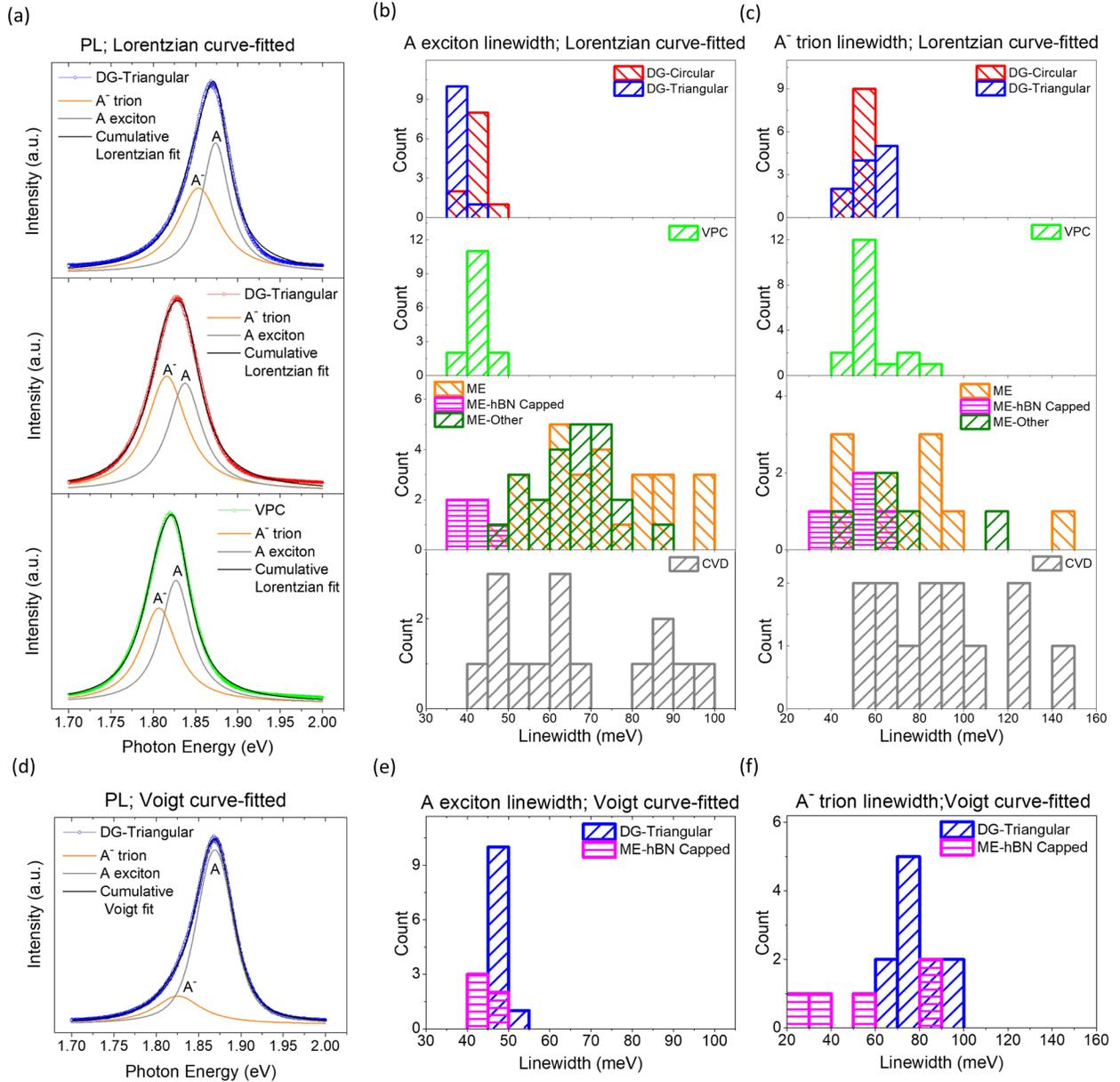

*Figure 4* *(a). Normalized PL vs. photon energy of directly grown triangular, circular and VPC-grown samples, with the Lorentzian curves fitted to the exciton and trion, the sum of which gives rise to the cumulative fit that is in agreement with the underlying PL spectra. (b) The histogram of A-exciton linewidths and (c) histogram of corresponding A⁻ trions linewidths -obtained from Lorentzian fits- of the 2D-MoS$_2$ samples fabricated by various techniques. All samples are either grown on Si/SiO$_2$ or transferred onto it. We have separated three different types of mechanically exfoliated samples, in which, ME means mechanically exfoliated uncapped, ME-hBN capped is evident by its name, and ME-Other means mechanically exfoliated samples that are either capped by other materials than h-BN or are post-processed with other methods to increase the lineshape quality. (d) Normalized PL vs. photon energy of directly grown triangular with the Lorentzian curves fitted to the exciton and trion, the superimpose of which gives rise to the cumulative fit, that is very well in agreement with the underlying PL spectra. (e) The histogram of A-exciton linewidths and (f) histogram of corresponding A⁻-trions linewidths -obtained from Voigt fits- of the 2D-MoS$_2$ directly grown triangles and ME-hBN capped samples.*

To perform the most stringent study of our samples and compare them with the best available other samples, *i.e.*, mechanically exfoliated h-BN-capped samples, we also fitted a set of Voigt functions to the triangular samples grown by our method as well as the h-BN-capped ME samples. Figure 4(d) shows a typical Voigt function fit to a PL spectrum obtained from our DG-triangular samples, elucidating the high quality of this fit. The results of these fitting analyses are shown in Figures 4(e), and (f). We find that even after using a Voigt fit, the linewidth qualities of our samples are well comparable to the h-BN-capped ME samples with 40-50 meV linewidth of A-exciton. For the A$^-$-trion, although the best h-BN-capped ME samples appear to have lower linewidths, the median values of the two types were comparable as well. Our detailed, systematic analysis shows that non-contact samples (DG-triangular) of 2D-MoS$_2$ synthesized using direct growth technique indeed results in high sample quality, with narrower room temperature A-exciton linewidths compared to all other known (unprocessed) methods, and closely comparable to h-BN encapsulated ME samples. Taken together with the simplicity of this approach, we believe this is a big step forward towards low-cost, high-quality and easily accessible technology for 2D material synthesis.

## Conclusion

Conventional methods of fabricating 2D-TMD devices all have limitations that make them challenging for practical use. While ME affords samples of high quality, it is not practical for fabricating 2D samples in large quantities. CVD synthesis provides scalability for practical application, but their material quality is still not electronic/optoelectronic grade. Based on PL and mobility measurements, the samples produced by VPC, our previously developed method, are superior in quality to CVD samples and the technique is scalable. However, the need for precise multi-parameter control makes it often challenging to get reproducible samples in a typical scientific laboratory – and this process is not amenable for on-demand miniaturization. In the current work, we show that it is possible to obtain high-quality 2D-MoS$_2$ ideal for various optoelectronic applications comparable to state-of-the-art MoS$_2$ samples, using a low-cost, flowless, facile single-pot method that circumvents the need for any chemical reactions. Our detailed PL and Raman analysis, especially the excitonic linewidth analysis results establish that in contrast with the common misconception, high-quality optoelectronic-grade 2D-$MoS_2$ can be acquired by methods as simple as direct growth by heating of bulk sources without the need for flowing carrier gases. The A-exciton linewidth of a triangular 2D monolayer crystal grown by our direct method, without the need for capping or annealing, is about ~ 35-40 meV, which is as sharp as the best attainable h-BN-capped ME samples; the A$^-$ trion linewidth is also quite sharp for our samples. Furthermore, our comprehensive linewidth analysis also indicates our direct method has far more sample-to-sample homogeneity, compared to other methods, including ME. It is quite remarkable that our approach, which in some sense is the simplest conceivable one for growing 2D materials, results in samples that have much narrower linewidths compared to those of as-exfoliated ME samples, and compare with ultra-flat h-BN-encapsulated ME samples, which have so far remained a Hallmark of 2D-TMD quality. We also show that by controlling the substrate's distance from the source, it is possible to obtain strained samples that have spatial defects created by strain-induced wrinkles on the grown 2D materials. As for the growth technique itself, *on one hand,* by overcoming the necessity of flowing a carrier gas, mass-flow-controllers, and multiple precursors, makes our method amenable for miniaturization since the confinement volume of Ar chamber can be suitably reduced to accommodate just the source and the substrate, and further allows the possibility of reducing the size of the furnace chamber, or non-standard approaches such as solar heaters. On the other hand, this novel technique is also amenable for scaled-up fabrication of 2D-TMDs on large-scale substrates. Other synthesis techniques have the limitations such as the need for a uniform carrier gas flow-rate on the surface of the substrate and a detailed control of distance/proportion of chemical precursors over a large area that makes fabricating 2D-TMDs on large-scale substrates almost impossible; this is where our direct growth technique has a novel advantage. Furthermore, alternative heating solutions, such as focused solar heating, are in principle compatible with our sealed tube method.

## Methods

**Source preparation.** As purchased $MoS_2$ powder can be used as the source, but we pre-processed the powder with a widely used technique named LPE – ultrasonication-assisted liquid phase exfoliation[157] – to acquire smaller flakes to increase the quality of the 2D samples. MoS$_2$ powder (99% Sigma Aldrich) was first dispersed in isopropyl alcohol (IPA) (99.5% Alpha Aesar) with the ratio of 1:10 and kept for 1 hour; then, the dispersion was ultrasonicated (UP100H Hielscher ultrasonic processor) 30 kHz and 80% of the power for 8 hours while the dispersion beaker was placed in room temperature water to avoid overheating. When finished, the top half of the suspension, which now contained 2D flakes of MoS$_2$ floating in the IPA, was collected and centrifuged for 2 min at 1000 rpm (Thermo Scientific centrifuge). The entire process was performed under ambient conditions (see Hejazi *et al*. for more details on the LPE technique). Afterward, the top half of the suspension was collected and used for 2D-MoS$_2$ growth.

**Direct growth of 2D-MoS$_2$.** We used a piece of Si chip instead of the crucible (*i.e.*, chip-crucible) and put a few drops of the above-described MoS$_2$ suspension on it; the IPA dries out in a few second, leaving small flakes of MoS$_2$ that can only be seen under a microscope (see Hejazi *et al*.). A 0.5 cm by 0.5 cm chip was cut from Si/SiO$_2$, *i.e.*, Si wafer with 300 nm of SiO$_2$ coating (Addison Wafer), and used as substrate. The chip surface was cleaned using a compressed air gun to remove dust, etc.; to obtain an even cleaner substrate, one can use IPA and deionized water before blowing the air. The substrate was then placed facing down directly on the chip-crucible, making a sandwich (see Figure 1(a)). Afterward, we placed the sandwich inside an alumina boat and slid the boat inside a quartz tube (AdValue technology). We examined two slightly different approaches.

In the first approach, we pumped the air out of the tube, filled it with Argon (99.99 Medical Technical Gases) up to atmospheric pressure (~ 760 Torr), and sealed it. The tube was heated from room temperature up to 650 °C at a rate of 100 °C/min; then it was heated to 750 °C at 5 °C/min. Throughout the heating process, pressure builds up inside the tube; in order to reduce the pressure, very slowly we opened the seal and let the extra Argon leave the tube once in every few minutes; we were observant of the tube pressure to make sure it did not drop below atmospheric pressure, to prevent re-entering the air into the tube which could cause contamination and compromise the growth. During the growth, the pressure was about 770-800 Torr. In the other approach, after pumping the air out and filling with Argon, we just filled it up to a fraction of atmospheric pressure, so when heating, even though the pressure builds up, it would not go much beyond the atmospheric pressure. The heating up step was the same for both approaches. After reaching 750 °C, we kept it for 40 min, and when finished, we opened the furnace and cooled down the tube as fast as possible using an air fan. The tube was kept sealed till it cooled down as low as room temperature. Afterward, the substrate was collected and used for optical measurements. We present the samples fabricated by the first approach since they possess better qualities,

**Optical measurements.** The PL and Raman spectra were collected using a Modu-Laser Stellar-ReniShaw Raman spectroscopy tool equipped with a 150 mW, 488 nm laser. The laser light was focused on 2D-$MoS_2$ single crystals for about 1 min in Raman-mapping and about 3 min in PL-mapping.

**Atomic Force Microscopy.** The AFM images of the 2D-$MoS_2$ samples were collected using FastScan AFM instrument (Bruker Instruments, Billerica, MA) at the FastScan' ScanAsyst Mode using ScanAsyst cantilevers (Bruker Instruments).

**Curve-fitting, and digitizing the PL images.** We used OriginPro commercial software to curve-fit the PL data. The PL data for our samples were available, but for comparison with previously published literature, if their PL data was not publicly available, we used the same software to import images from the articles and digitized them to extract the PL data. When fitting a function, we used OriginPro to fit superposition of either two Gaussians, two Lorentzians or two Voigt packages, one for A-exciton and one for A⁻ trion. In probability theory, a normal (or Gaussian or Gauss or Laplace–Gauss) distribution is a type of continuous probability distribution for a real-valued random variable. The general form of its probability density function is

$$G(x; \mu, \sigma) = \frac{1}{\sqrt{2\pi\sigma^2}} e^{-\frac{1}{2}\left(\frac{x-\mu}{\sigma}\right)^2}$$

The parameter $\mu$ is the mean or expectation of the distribution (and also its median and mode), while the parameter $\sigma$ is its standard deviation. The variance of the distribution is $\sigma^2$.

Lorentz distribution, also known as the Cauchy distribution, Lorentzian function, Cauchy–Lorentz distribution, or Breit–Wigner distribution is also a continuous probability distribution, and the general form of its probability density function is

$$L(x; x_0, \gamma) = \frac{1}{\pi \gamma \left[1 + \left(\frac{x - x_0}{\gamma}\right)^2\right]}$$

where $x_0$ is the location parameter, specifying the location of the peak of the distribution, and $\gamma$ is the scale parameter which specifies the half-width at half-maximum (HWHM), alternatively $2\gamma$ is full width at half maximum (FWHM).

The Voigt profile (named after Woldemar Voigt) is a probability distribution given by a convolution of a Cauchy-Lorentz distribution and a Gaussian distribution. It is often used in analyzing data from spectroscopy or diffraction. Without loss of generality, we can consider only centered profiles, which peak at zero. The Voigt profile is then

$$V(x; \sigma, \gamma) = \int_{-\infty}^{\infty} G(x'; \sigma) L(x - x'; \gamma) \, dx'$$

where $x$ is the shift from the line center, $G(x; \sigma)$ is the centered Gaussian profile ($\mu = 0$), and $L(x; \gamma)$ is the centered Lorentzian profile ($x_0 = 0$). As we can see, the Voigt function does not have a closed-form solution.

**Conflict of Interest:** The authors declare no competing financial interest.

## Acknowledgment

The authors DH, RT, and SK would like to gratefully acknowledge financial support from NSF ECCS 1351424, and a Northeastern University Provost's Tier 1 Interdisciplinary seed grant, and a Northeastern University GapFund 360 Phase I award. SK also acknowledges Indo-US grant, IUSSTF/JC-071/2017.

## Author Information

**ORCID**

Davoud Hejazi: 0000-0002-5215-6395

Renda Tan:

Neda Kari Rezapour: 0000-0002-3528-8193

Mehrnaz Mojtabavi: 0000-0002-0483-3582

Meni Wanunu: 0000-0002-9837-0004

Swastik Kar: 0000-0001-6478-7082